
\tolerance=1200
\magnification 1200
\baselineskip=12pt plus 1pt
\parindent=25pt

\font\small=cmr10 at 10truept
\baselineskip=20pt plus 1pt


{\nopagenumbers
{
\small
\baselineskip=12pt plus 1pt
\hfill Alberta-THY-8-1993

\hfill FEBRUARY 1993
}

\vskip.7cm
{\bf
\centerline{Miens of The Three Dimensional Black Hole}
}

\vskip.7cm

\centerline{{\bf Nemanja Kaloper}$~^{*}$}
{
\small
\baselineskip=12pt plus 1pt
\footnote{}{$^{*}~~$email: kaloper@fermi.phys.ualberta.ca }
}
\centerline{Theoretical Physics Institute}
\centerline{Department of Physics, University of Alberta}
\centerline{Edmonton, Alberta T6G 2J1, Canada }
\vskip.5cm
\centerline{\bf Abstract}

{
Some aspects of the rotating three-dimensional
Einstein-Anti-de-Sitter black hole
solution, constructed recently by Banados, Teitelboim and Zanelli are
discussed. It is shown explicitly that this black hole
represents the most general black hole type solution of the
Einstein-Anti-de-Sitter theory. The interpretation of one of the
integrals of motion as the spin is discussed.
Its physics relies on the topological structure
of the black hole manifold, and the notion of simultaneity
of space-like separated intervals. The relationship of the black hole solution
to string theory on a $2 + 1$ dimensional target space is examined, and
it is shown that the black hole can be understood as a part of the full
axion-dilaton-gravity, realized as a WZWN $\sigma $ model.
In conclusion, the pertinence of this solution to
four-dimensional black strings and topologically massive gravity is pointed
out.
}
\vskip1cm
\centerline{\it Submitted to Phys. Rev. {\bf D}}
\vfil
\eject}
{\bf 1.~~Introduction}
\vskip0.5cm

The black hole conundrum has long been one of the most outstanding
problems of modern physics. It has remained in focus as one of
the potential testing grounds for quantum-gravitational phenomena
for a long time.
The formal difficulties of four dimensional
gravity, however,
have often made the study of black holes inherently more
complicated. To surmount some of these
difficulties many researchers have resorted to the models of
gravity in dimensions lower than four, in hope that the essential properties
of black holes in lower dimensions will model reasonably accurately
those of the four dimensional solutions.
One such attempt has resulted recently in the
construction of the Einstein-Anti-de-Sitter rotating black hole
in three dimensions by Banados, Teitelboim and Zanelli (hereafter refered to as
BTZ)[1]. Their solution  has attracted further
attention as it has later been shown how
it can be obtained by restricting a four dimensional
Minkowski manifold of signature zero on a coset [2-3], followed
by the one-point compactification of one of the coordinates to a circle.
Furthermore, the conditions under which such black holes can form in
a collapse of matter in conventional General Relativity
have been investigated in [4].
The purpose of this work is to demonstrate
that their solution can be easily incorporated in
the framework of string theory with some minor extensions [5].
Namely, the BTZ solution with half the initial cosmological constant can
be extended with the inclusion of the antisymmetric Kalb-Ramond axion field
carrying the other half of the cosmological constant, and then reinterpreted
as either an ungauged or extremely gauged Wess-Zumino-Witten-Novikov (WZWN)
$\sigma$ model derived from the group $SL(2,R)$ [6-11]. The one-point
compactification of one of the coordinates can be accomplished either
by factoring out a discrete group in the ungaguged construction, or requiring
that the model lives on a coset $SL(2,R) \times R/ R$.

This paper is organized as follows. In section 2., I will derive
the solution from solving the equations of motion,
by employing the Kaluza-Klein
dimensional reduction from three dimensions to one [12], and show that the
solution of Ref. [1] is the unique solution of three-dimensional Einstein
gravity with a negative cosmological constant which features horizons.
In section 3. I will comment on the interpretation of one
of the constants of motion as the spin of the black hole, and show that this
stems from interweaving the topological structure of the manifold with the
requirement of global simultaneity of space-like intervals. Section 4.
concentrates on the stringy interpretation of the solution and demonstrates
how the solution is realized as a WZWN $\sigma$ model. Lastly, I will
comment on the relationship of this solution to topologically massive gravity
[13], and cosmic strings.

\vskip1cm
{\bf 2.~~ Classical Theory}
\vskip0.5cm

The classical theory is defined with the Einstein-Hilbert action in three
dimensions,
$$
S=\int d^{3}x\sqrt{g}~\big({1 \over 2\kappa^{2}}R + \Lambda \big) \eqno(1)
$$

\noindent where $R$ is the Ricci scalar and $\Lambda$ the cosmological
constant. The conventions employed here are that the metric is of signature
$+2$, the Riemann tensor is defined according to
$R^{\mu}{}_{\nu\lambda\sigma} = \partial_{\lambda}\Gamma^{\mu}{}_{\nu\sigma} -
\ldots $, and the cosmological constant is defined with the opposite sign
from the more usual conventions: here, $\Lambda > 0$ denotes a negative
cosmological constant. In the remainder of this paper, I will
work in the Planck mass units: $\kappa^2 = 1$.

The Einstein equations associated with this theory, in the absence of other
sources yield the locally trivial solution:
$R^{\mu}{}_{\nu\lambda\sigma} =
-\Lambda (\delta^{\mu}_{\lambda} g_{\nu\sigma}
- \delta^{\mu}_{\sigma} g_{\nu\lambda})$ which suggests that the unique
solution is the Anti-de-Sitter space in three dimensions. However, there
appear nontrivial configurations in association with the global structure of
the manifold described with the above curvature tensor. It is interesting
to note that all the metric solutions have well defined
curvature, except possibly at a point later to be identified with the
black hole singularity [2-3].

Therefore, to inspect all the possible solutions one should resort to a
closer scrutiny of the problem at hand. The investigation of [1-3]
demonstrates how nontrivial black hole solutions can be obtained by
factorization and topological identification in the Anti-de-Sitter
manifold. However, a particularly simple procedure can be followed, where one
solves the differential equations derived from (1) and investigates allowed
values for the integration constants. In addition, this procedure yields
further information regarding whether all possibilities for the construction
of nontrivial solutions are exhausted by the above mentioned identifications.

Instead of writing out explicitly the Einstein's equations, I will here
work in the action, as this approach offers an especially simple way to find
the solutions. The background ans\" atz is that of a stationary
axially symmetric metric:
$$
ds^{2}= \mu^2 ~dr^{2}+G_{jk}(r)~dx^j dx^k
\eqno(2)
$$

\noindent where the $2 \times 2$ matrix $G_{jk}(r)$ is of signature $0$ as
the metric (2) is Lorentzian and one of the coordinates $\{x^k\}$ is timelike.
The ``lapse'' function $\mu^2$ is kept arbitrary as its variation in (1) yields
the constraint equation. The cross terms $drdx^k$ corresponding to the
``shift'' functions can be removed by coordinate transformations
$x^k \rightarrow x^k + F^k(r)$.

The metric above clearly has two toroidal coordinates $\{x^k\}$ which are
dynamically unessential. Hence the problem is effectively one-dimensional. The
Kaluza-Klein reduction, with rescaling of the action (1) according to
$S_{\rm eff} = 2 S/\int d^2x$ yields
$$
S_{\rm eff}=\int dr ~\mu e^{-\phi}\big({1 \over \mu^{2}}\phi'^2 + 2 \Lambda
+ {1 \over 4 \mu^{2}}~ Tr G'^{-1} G' \big) \eqno(3)
$$

\noindent with the ``dilaton'' field $\phi$ being constrained (rather, defined)
by $\exp(-2\phi) = - \det G$. The minus sign here follows from the fact that
${\rm sign}(G) = 0$, i.e., $\det G < 0$. The prime denotes derivative
with respect to $r$. Thence, the problem is reducible to a simple
mechanical system describing ``motion'' of the matrix $G$ with
several rheonomic solvable constraints. As such, the ``dilaton'' constraint
above can be solved for $\phi$, which thence may be completely eliminated from
the action. However, it is instructive to keep the explicit dilaton in (3) and
enforce the above constraint with help of an additional Lagrange multiplier
$\lambda$. Furthermore, there is an additional simplification coming from
properties of $2 \times 2$ matrices. In the above equations for the action and
the ``dilaton'' the inverse and determinant of $G$ figure explicitly, thus
giving the problem in question the appearance of a highly non-linear one.
The $2 \times 2$ magic comes to the rescue: it is possible to
reexpress the action in Gaussian form in terms of $G$ only. From
$$\eqalign{
&\det G = {1 \over 2} Tr \bigl(\epsilon G\bigr)^2 \cr
&~G^{-1} = {1 \over \det G}\epsilon G \epsilon }
\eqno(4)
$$
\noindent with $\epsilon = {\rm i} \sigma_2$ being
the two dimensional antisymmetric
symbol, the action (3) can be rewritten as
$$
S_{\rm eff}=\int dr ~\bigl\{2 \Lambda \mu e^{-\phi}
+ {e^{\phi}\over 4 \mu} \bigl[Tr \bigl(\epsilon G'\bigr)^2 + \lambda
{}~Tr \bigl(\epsilon G\bigr)^2 \bigr] - {\lambda e^{-\phi}\over 2 \mu}
\bigr\} \eqno(5)
$$

\noindent Clearly, the theory has three Lagrange multipliers: $\mu$, $\phi$
and $\lambda$, which all propagate according to algebraic equations. Thus
the associated equations of motion are very simple. Indeed, the standard
variational procedure leads to
$$\eqalign{
&2 \Lambda \mu e^{-\phi} + {e^{\phi}\over 4 \mu}
Tr \bigl(\epsilon G'\bigr)^2 = 0\cr
&~~{e^{\phi}\over 4 \mu} Tr \bigl(\epsilon G'\bigr)^2 +
{e^{-\phi}\over 2 \mu} = 0\cr
&~~~~~~~~~{ \lambda e^{-\phi}\over 2 \mu} = 0\cr
&~~~~~\bigl( {e^{\phi}\over \mu} G' \bigr)' =
{ \lambda e^{\phi}\over \mu} G\cr}
\eqno(6)
$$

\noindent Obviously, $\lambda = 0$. The system of equations above simplifies
to
$$\eqalign{
&2 \Lambda \mu e^{-\phi} + {e^{\phi}\over 4 \mu}
Tr \bigl(\epsilon G'\bigr)^2 = 0\cr
& ~~~Tr \bigl(\epsilon G'\bigr)^2 +
2e^{-2\phi} = 0\cr
&~~~~~~\bigl( {e^{\phi}\over \mu} G' \bigr)' = 0 \cr}
\eqno(7)
$$

\noindent and with using the gauge freedom expressed by the arbitrary
``lapse'' $\mu$ and fixing the gauge to $\mu \exp{(-\phi)} = 1$, the solution
is easy to find. It is just
$$\eqalign{
&\mu^2 = -{1 \over \det G} \cr
&~G = Cr + D \cr}
\eqno(8)
$$

\noindent and $C,D$ are constant symmetric matrices determined from the initial
conditions, and the constraint $ \det C = - 4 \Lambda$. The minus sign in (8)
is precisely the same one discussed following Eq. (3).

What remains is to  analyse the values of the integration constants. To begin
with, the metric can be rewritten as
$$
ds^{2}= - { dr^{2} \over \det(Cr + D)} + (Cr + D)_{jk}~dx^j dx^k
\eqno(9)
$$

\noindent Since $C$ is symmetric and nonsingular, ($\det C = -4\Lambda \ne 0$),
it can be diagonalized with an orthogonal transformation. So, $C =
O^{T}C_{d}O$. From the metric (9) such transformation is just a coordinate
transformation of the $\{x^k\}$ part of the metric,
$x^k \rightarrow O^k{}_{j}x^j$. Thence $C$ could have been assumed diagonal
from the beginning. Furthermore, its eigenvalues $c_1,c_2$ can be set equal to
$\pm 1$ by a scale transformation $x^k \rightarrow x^k / \mid c_k \mid$. Thus
$C$ is just the $1 + 1$ Minkowski metric, $C = \eta = {\rm diag}(1, -1)$. At
this point one could object that the rescaling can introduce nontrivial deficit
angle if a coordinate $x^k$ is compact. This can be restored later
by changing the period of compactification. Moreover, the diagonalization
of $C$ can also be accomplished with a shift of the spacelike coordinate by a
linear function of time. Therefore, the above discussion is fully justified.

The next step is the matrix $D$, which only has to be symmetric. None of the
above manipulations with coordinates in order to reduce $C$ to the $1 + 1$
Minkowski metric affects the general structure of the matrix $D$. Therefore,
the $2\times 2$ metric can be written as
$$
G = \eta r + D = \left(\matrix{r + d_{11} & d_{12}\cr
 d_{12} & -r + d_{22} \cr}\right)
\eqno(10)
$$

\noindent and evidently, one of the diagonal elements of $D$
can be removed by a shift in $r$.
This indicates an additional requirement which ought to be
imposed on $D$. Since one is interested in a black hole type solution, with
physical horizons defined as the hypersurfaces where the timelike Killing
vector outside the black hole has vanishing norm, and flips into spacelike
after passing through the horizon, it must be $d_{22} \ge - d_{11}$. Then,
one can simply set $d_{11}=0$. Lastly, if the
spacelike coordinate $\theta$ is to be interpreted
as an angle, the identification $\theta \cong \theta + 2\pi$ must
be made. Thus, the final solution is
$$
ds^{2}=  { dr^{2} \over 4 \Lambda [r(r - d_{22}) + d^2_{12}]} +
{}~~(d\theta, dt) \left(\matrix{r & d_{12}\cr
 d_{12} & -r + d_{22} \cr}\right) \left(\matrix{d\theta\cr
 dt\cr}\right)
\eqno(11)
$$

The Eq. (11) is precisely the solution of Ref. [1] as can be seen after a
coordinate transformation. The integration constants can be rearranged by
introducing the mass $M = d_{22}\sqrt{\Lambda}$ and the spin
$J = - 2 d_{12} \sqrt{\Lambda}$, as well as the parameter measuring the
position of the horizon in the new coordinates:
$\rho_+^2 = M(1 - (J/M)^2)^{1/2}$.
With the definitions
$R^2 = r = (\sqrt{\Lambda}/2)\bigl(\rho^2 + M - \rho_+^2 \bigr)$ and
$N^{\theta} = - J /2R^2$ the metric (11) can be put in the
BTZ form:
$$
ds^{2}=  { d\rho^{2} \over \Lambda (\rho^2 - \rho^2_{+})} +
{}~R^2 (d\theta + N^{\theta} dt)^2
- {\rho^2 \over R^2} {\rho^2 - \rho^2_{+} \over \Lambda} dt^2
\eqno(12)
$$

\noindent From the formulas above one finds that physical black holes should
also satisfy the constraint $\mid J \mid \le M$.
If this were not fulfilled, one would end
up with a singular structure, manifest by the appearance of closed timelike
curves in the manifold accessible to an external observer, crossing the
point $R =0$. Such a voyage has been investigated in [5] for the spinless
case, and also in [9] for the vacuum. Moreover, it has been
argued that, although the solution (12) does not have curvature singularities,
they can develop if the metric is slightly perturbed by
a matter distribution [2-3].
Thus, the singularities are hidden by a horizon if the spin is bounded above
by the mass. Thermodynamics of (12) has been analysed in [1-2], where the
Hawking temperature has been calculated.
The solution with $J=M$ is understood as the extremal black hole, and $J=M=0$
serves the
role of the vacuum. These two solutions actually appear to have similar
local properties, as will be discussed in the next section.
The Anti-de-Sitter metric is recovered with
$J=0, M=-1$ [1-3].

\vskip1cm
{\bf 3.~~ Spin And Simultaneity}
\vskip0.5cm

There still remains to determine the physical nature of the spin $J$.
It has been so interpreted by the careful examination of the boundary
terms in the action which appear in the ADM formulation of General Relativity
in three dimensions [1-2].
Yet an interesting observation is in place here. The $r$ dependent part
of the metric $G$ is an $SO(1,1)$ invariant, being the $1 + 1$ Minkowski
metric. Then one can ask if the matrix $D$ can be diagonalized by a coordinate
transformation. Indeed, the transformation
$x'^k = \tilde O^k{}_{j}x^j$ where
$$
 \tilde O = \left(\matrix{~\cosh \beta~&~\sinh \beta~\cr
\sinh \beta~&~\cosh \beta~\cr}\right)
\eqno(13)
$$

\noindent and
$$
\sinh \beta = {\rm sign}(J) {1 \over \sqrt{2}} \Bigl(
{1 - \sqrt{ 1 - (J/M)^2} \over \sqrt{ 1 - (J/M)^2} }\Bigr)^{1/2}
\eqno(14)
$$

\noindent removes the cross term $d\theta dt$ from the metric, and is
clearly valid for all physical black holes with $\mid J \mid \le M$.
In terms of the new coordinates the metric (12) can be rewritten as
$$
ds^{2}=  { d\rho^{2} \over \Lambda (\rho^2 - \rho^2_{+})} +
{}~\rho^2 d\theta'^2
- {\rho^2 - \rho^2_{+} \over \Lambda} dt'^2
\eqno(15)
$$

\noindent This solution describes a black hole of spin $J'=0$ and mass
$M' = M \bigl( 1 - (J/M)^2\bigr)^{1/2}$.

The question
one should ask is, if the transformation (13) is globally defined. If the
answer is positive, then the angular momentum in the metric would be spurious.
What can be seen immediately is that with the
help of (13), which corresponds to
a ``boost'' in the azimuthal direction, a comoving observer can be found
who will not be able to discern the influence of the angular momentum by any
local experiment. Hence (12) and (15) are completely equivalent locally.

The answer is that due to the identification
$\theta \cong \theta + 2\pi$ the global structure of the manifold
with the metric (12) is not invariant under a coordinate transformation
generated by a ``boost'' (13). This can be seen as follows.
The manifold can be foliated by cylinders
$R \times S^1$ corresponding to constant $r$ (or $\rho$). The cylinders
in the frame where the identification has been made (and the spin $J$ has been
defined) can be represented as rectangular patches in the $t-\theta$ plane with
edges at $\theta = 0$ and $\theta = 2 \pi$ identified along the congruence
$t = 0$. After the boost has been performed, in the new coordinates the
manifold
is represented with patches tipped with respect to the $t'$ axis by angle
$\cos^{-1}\cosh \beta$ and identification
now goes along the congruence
$t' = \cosh \beta ~\theta'$. Hence, the global simultaneity of spacelike events
is lost! If one goes around the universe, with a clock which remembers
the initial point, upon the return to it the clock reading exibits a discrete
jump. Therefore, to measure the spin of the
black hole, one can build a stroboscopic device by measuring the discrepancy
of the arrival time of light rays sent around the black hole in opposite
directions. Boosting in the azimuthal direction, observers can bring themselves
to the frame where the light signals arrive back simultaneously, and then
measure the spin. It is interesting to note that precisely the same phenomenon
can be found in Special Relativity on a cylinder with flat metric.

The discussion above gives an interesting connection between the vacuum
$J=M=0$ and an extremal black hole $J=M$. If one takes the limit
$M \rightarrow J + 0^+$ in the boosted coordinates (15), the metric reduces to
the vacuum solution $J' = M' = 0$, but the coordinate transformation (13) is
ill-defined in the limit, since $\tanh \beta \rightarrow 1$. But this
corresponds to boosting up to the speed of light in Special Relativity,
where in the $1 + 1$ case, the space-time tends to a degenerate case for such
an observer. Thus one can think of the extremal black hole as the maximally
boosted vacuum, up to the global structure.

The physical interpretation of spin therefore derives from the global
properties of the manifold. Essentially, the spin is introduced by
choosing a special observer who is granted the judgement how to perform
the identification.
More elaborate, but similar properties have been
found by Misner in the connection with Taub-NUT spacetimes [14]. These
conclusions are in prefect agreement with the constructions of Refs [1-3].

\vskip1cm
{\bf 4.~~ Wess-Zumino-Witten-Novikov $\sigma$ Model Approach}
\vskip0.5cm

In this section I will discuss the relationship of the solution (12)
to string theory and show how it can be extended to represent an exact string
solution too. In order to do it, an elementary review of
the WZWN $\sigma$ model
approach is provided first.
The dynamics of string theory on the world-sheet
is defined by the tree-level Polyakov action
$$
S_{\sigma}~={1 \over \pi}
{}~\int d^2\sigma (G_{\mu\nu} + 2\sqrt{2 \over 3}
B_{\mu\nu})~\partial_{+}X^{\mu}
\partial_{-}X^{\nu}
\eqno(16)
$$

\noindent  where $~G_{\mu\nu}~$ and $~B_{\mu\nu}~$ are
the world sheet target
metric and the Kalb-Ramond antisymmetric field. The
rather unusual factor $~2\sqrt{2/3}~$ in Eq. (1) is
introduced following the normalization convention
in earlier work, where the wedge product
of two forms is defined by $~\alpha \wedge \beta =
Alt~(\alpha \otimes \beta)$ as opposed to the other usual convention,
$~\alpha \wedge \beta = {(p + q)! \over p! q!}
Alt~(\alpha \otimes \beta)$. The action (16) in general also includes the
dilaton, but it can be computed in the semiclassical approach from the
associated effective field theory on target space.
Its effective action is,
in the world sheet frame and to order $O(\alpha'^{0})$,
$$
S~=~\int d^{3}x\sqrt{G}  e^{-\sqrt{2}\kappa \Phi}
\big({1 \over 2\kappa^{2}}R -
H_{\mu\nu\lambda}H^{\mu\nu\lambda}+
\partial_{\mu}\Phi \partial^{\mu} \Phi + \Lambda \big) \eqno(17)
$$

\noindent Here
$~H_{\mu\nu\lambda}=\partial_{[\lambda}B_{\mu\nu]}~$
is the field strength associated with the Kalb-Ramond
field $~B_{\mu\nu}~$ and $~\Phi~$
is the dilaton field, which appears naturally in the string sector
and whose dynamics guarantee the conformal anomaly cancellation.
Braces denote antisymmetrization over enclosed indices.
The cosmological constant has been included to represent the
central charge deficit
$~\Lambda={2 \over 3}\delta c_{T} = {2 \over 3}(c_{T}-3) \ge 0~$.
It arises
as the difference of the
internal theory central charge and the total central charge for a
conformally invariant theory $~ c_{tot}=26~$ [6-11,15].

The WZWN approach starts with the construction of the field theory
on the world-sheet defined by the WZWN $\sigma$ model action
on level $~k~$,
$$
S_{\sigma}~=~{k \over 4 \pi} \int d^2\sigma
Tr\bigl(g^{-1}\partial_{+}g\,g^{-1}\partial_{-}g\Bigr)
- {k \over 12 \pi}~\int_{M} d^3 \zeta
Tr\Bigl( g^{-1}dg \wedge g^{-1}dg \wedge g^{-1}dg \Bigr)
\eqno(18)
$$

\noindent where $g$ is an element of some group $G$. The action above has
a very big global invariance, the continuous part of which is $G\times G$.
One way to construct the string solutions of this
theory, which can be put in form (16) is
choosing a group $G$, the parameter space of which
represents the target manifold, and
maintaining conformal invariance. The other may be to identify a part of the
parameter manifold by factoring out locally a subgroup of the
global invariance group $G \times G$. This is accomplished with
choosing an anomaly-free subgroup $H \subset G \times G$ and gauging it
with stationary gauge fields.
Either way, after the group has been parametrized,
(18) can be rewritten in terms of the parameters in the
form (16) and the metric
and the axion are just simply read off from the resulting expressions. The
dilaton then can be computed from the effective action (17), as has been
mentioned above. Its appearance owes to the requirement of conformal
invariance. In the remainder of this section I will demonstrate how the
solution (12) arises in this approach as the gravitational sector of the
WZWN constructions in two different ways.

I will first demonstrate that the theory described by (18) with the group
$G = SL(2,R)/P$ contains the solution (12). The discrete group $P$ will be
specified later. The central charge
of the target for this model for level $~k~$ is
$~c_{T}=3k / (k-2)~$. Thus the central charge deficit, by the formulae above,
will be given by
$$
\delta c_{T}~=~{6 \over k - 2}~\simeq ~{6 \over k}
\eqno(19)
$$

\noindent in the semiclassical limit $~k \rightarrow \infty~$,
where the theory is most reliable. Therefore, the cosmological constant
is $\Lambda = 4/k$.
The group $SL(2,R)$
can be parametrized according to
$$
 g~=~\left(\matrix{~e^{\sqrt{2 \over k}q\theta'} \cosh \vartheta~
 &~e^{\sqrt{2 \over k} qt'} \sinh \vartheta~\cr
{}~e^{-\sqrt{2 \over k} qt'} \sinh \vartheta~
 &~e^{-\sqrt{2 \over k}q\theta'} \cosh \vartheta~\cr }\right)
\eqno(20)
$$

\noindent where $q$ is an arbitrary constant.
In terms of these parameters, the action (18) can be rewritten as
$$\eqalign{
S_{\sigma}~=~{1\over \pi} &\int d^2\sigma
\Bigl({k \over 2} \partial_{+}\vartheta \partial_{-}\vartheta
+ q^2 \cosh^2 \vartheta \partial_{+}\theta' \partial_{-}\theta'
- q^2 \sinh^2 \vartheta \partial_{+}t' \partial_{-}t' \Bigr) \cr
+ {\sqrt{2k} \over \pi} &\int d^2\sigma q \bigl(\sqrt{2 \over k} qt'
+ \ln \sinh \vartheta \bigr) \sinh \vartheta \cosh \vartheta
\Bigl(\partial_{+}\theta' \partial_{-}\vartheta
- \partial_{-}\theta' \partial_{+}\vartheta \Bigr)
 \cr}
\eqno(21)
$$

\noindent Comparing with (16), one deduces
$$\eqalign{
& ~~~~~~~~~~~~~~~~G_{\mu\nu}~=~\left(\matrix{{k \over 2}&~0~&~0~\cr
{}~0~&~q^2 \cosh^2 \vartheta~&~0~\cr
{}~0~&~0~&- q^2 \sinh^2 \vartheta \cr} \right) \cr
& B_{\mu\nu}~=~\sqrt{3k \over 4}q\bigl(\sqrt{2 \over k} qt'
+ \ln \sinh \vartheta \bigr) \sinh \vartheta \cosh \vartheta
\left(\matrix{~0~&~-1~&~0~\cr
{}~1~&~0~&~0\cr
{}~0~& ~0~&~0~\cr}\right) \cr}
\eqno(22)
$$

\noindent The metric $G_{\mu\nu}$ is exactly the canonical metric
on $SL(2,R)$, induced by the map of the Cartan-Killing form in the
neighborhood of unity in the Lie algebra $Tr\bigl( g^{-1} dg \bigr)^2$.
This is no surprise, since this is exactly the $\sigma$ model part
of the action. The axion is induced completely by the Wess-Zumino term.
The dilaton for this solution actually is constant, as can be readily verified
from the effective action (17). The dilaton equation of motion is
$$
\nabla^2 \Phi + \sqrt{2} \Bigl( {1 \over 2} R -
H^2 + \Lambda \Bigr) = 0
\eqno(23)
$$

\noindent and the substitution of the solution (22)  in (23), yields
$$
\Lambda~=~{Q^2 \over 3} e^{2\sqrt{2} \Phi_0} \eqno(24)
$$

\noindent where $Q$ is the $~3$-form cohomology charge of the
axion field $H=dB$, defined by the ``Gauss law'',
which, since ${^*}H$ is a zero form, is just
$Q~=~{1 \over 2 \pi} e^{-\sqrt{2} \Phi_0}~{^{*}H} = {\rm const}$. In order to
make contact with the solution (12), a change of coordinates and
the compactification of the spatial
coordinate $\theta'$ need to be made. To do this, I will show that the solution
(22) is equivalent to (11). Namely, the ``radial'' coordinate is introduced
by $r = q^2 \cosh^2 \vartheta$. Then, the metric can be rewritten as
$$
ds^{2}=  {1 \over 2 \Lambda}{ dr^{2} \over r(r - q^2)} +
{}~~r~d\theta'^2 -~ (r - q^2)~dt'^2
\eqno(25)
$$

\noindent This is almost precisely the solution (11), with $d_{12} =0$ and
$d_{22} = q^2$. The only difference is, the cosmological constant in  (11) is
half that in (25).  The reason for this discrepancy is, that the presence of
the axion introduces an extra contribution to the cosmological constant, which
just cancels one half of it, since the dilaton is constant. A careful
examination of duality transformations of the action (17) confirms this.
The axion field can be rewritten as
$
B =  \sqrt{3/8}~ \bigl(t' + \sqrt{k/2q^2} \ln\bigl((r/q^2) - 1\bigr) \bigr)
{}~d\theta' \wedge dr$, in form notation, and after the above
transformation of coordinates.
The axion is apparently time-dependent, which can be remedied with
recalling the gauge invariance of the axion: $B$ and $B' = B + d\Upsilon$ both
describe the same physics. Then, if
$\Upsilon = \sqrt{3/8}
\Bigl\{t'r + \sqrt{k/2q^2}(r - q^2)
\bigl[\ln\bigl((r/q^2)^2 - 1\bigr) - 1\bigr]\Bigr\}d\theta'$,
the gauge transformed axion
is
$$
B =  - \sqrt{3 \over 8}~ r ~d\theta' \wedge dt'
\eqno(26)
$$

\noindent The solution (25-26) has already been found previously, in Ref. [5].
At this point, one can perform the identification. Normally, it can be
accomplished by identifying $\theta' \cong \theta' + 2\pi$. This
would correspond to factoring out the discrete group $P' = \exp(2n\pi \xi')$,
with $\xi' = \partial/\partial \theta'$ the translational Killing vector
generating motions along $\theta'$ and $n$ integers, from $SL(2,R)$. The
resulting metric would have zero angular momentum, $J=0$. However, as was
discussed in Sec. 3., one can arbitrarily choose to identify along any boosted
translational Killing vector which is spatial outside the horizon. Hence, in
order to get the solution (12) with the mass $M$ and the angular momentum $J$,
one can boost back the coordinates $\theta', ~t'$ by (13) and identify points
by factoring the subgroup $P = \exp(2n\pi \xi)$ out of $SL(2,R)$, where
$\xi = \partial/\partial \theta
= \cosh \beta \partial/\partial \theta'
- \sinh \beta \partial/\partial t'$, and $\sinh \beta$ is given in Eq. (14).
The resulting configuration is the metric (12) extended with the axion (26).

There is a minor subtlety here. In order to complete the
identification, the axion solution has been gauge transformed  by a gauge
transformation which involves explicitly the compactified variables ($t'$) and
therefore is not single-valued on the covering space of the manifold. In this
respect, it can be treated as a ``large'' gauge transformation. Furthermore,
since the target has been compactified along $\theta$, in general when
a closed string
moves on such a world sheet there appear the winding modes associated with the
compact directions of the target [16].
The winding modes in principle can interact with the
axion field before and after the gauge transformation differently, and thus
distinguish between the solutions (22) and (25-26). This can be avoided if
one resorts to a different way of constructing the black hole solution (25-26).
The reason for the appearance of the gauge transformation $\Upsilon$ was
that the axion field has arisen from the Wess-Zumino term in (18). A simple
remedy is to gauge the WZWN $\sigma$ model on $SL(2,R) \times R$ by modding
out the axial vector subgroup of $SL(2,R) \times SL(2,R)$ mixed with
translations along $R$,
and taking the extremal limit where all the gauging is along $R$ [9].
This amounts to taking for the target the coset $SL(2,R) \times R / R$.
(A similar
procedure has been performed in [17], where a closed Bianchi I cosmology was
constructed.)
The central charge of this target for level $~k~$ is
$c_{T}=3k / (k-2) + 1 - 1$, where $\pm 1$ correspond to the
free boson and the gauging, respectively. Hence, $c_{T}=3k / (k-2)$ and
the cosmological constant is still $\Lambda = 4/k$.
The resulting solution is exactly (25-26), as can be easily verified.
The group $SL(2,R) \times R$ can now be parametrized as
$$
 g~=~\left(\matrix{~a~&~u~\cr
 -v~&~b~\cr}\right)~e^{{q \over \sqrt{k}}\theta'}
\eqno(27)
$$

\noindent where $~ab + uv = 1~$, the explicit form
of the ungauged sigma model of Eq. (18) is
$$\eqalign{
S_{\sigma}~=~&-{k \over 4 \pi} \int d^2\sigma
\Bigl(\partial_{+}u \partial_{-}v
+ \partial_{-}u \partial_{+}v + \partial_{+}a \partial_{-}b
 + \partial_{-}a \partial_{+}b \Bigr) \cr
&+ {k \over 2 \pi} \int d^2\sigma \ln{u}
\Bigl(\partial_{+}a \partial_{-}b
- \partial_{-}a \partial_{+}b \Bigr)
+ {q^2 \over 2\pi} \int d^2\sigma \partial_{+}\theta' \partial_{-}\theta' \cr}
\eqno(28)
$$

\noindent The gauge transformations corresponding
to the axial subgroup of $~SL(2,R) \times SL(2,R)~$ mixed with
translations along the free boson are
$$\eqalign{
&\delta a = 2\epsilon a ~~~~~\delta b = - 2\epsilon b
{}~~~~~\delta u = \delta v = 0 \cr
&~~~~~~~~~\delta \theta' = {2 \sqrt{2} \over q}\epsilon c
{}~~~~~~\delta A_{j} = -\partial_{j}
\epsilon \cr}
\eqno(29)
$$

\noindent and the gauged form of the sigma model (6) is
$$\eqalign{
S_{\sigma}(g, A)~=~S_{\sigma}(g)
&+ {k \over 2 \pi} \int d^2\sigma A_{+}
\Bigl( b\partial_{-}a - a\partial_{-}b
- u\partial_{-}v + v\partial_{-}u
+ {4 qc \over \sqrt{2}k} \partial_{-}\theta' \Bigr) \cr
&+ {k \over 2 \pi} \int d^2\sigma A_{-}
\Bigl( b\partial_{+}a - a\partial_{+}b
- v\partial_{+}u + u\partial_{+}v
+ {4 qc \over \sqrt{2}k} \partial_{+}\theta' \Bigr) \cr
&+ {k \over 2 \pi} \int d^2\sigma 4 A_{+}A_{-}
\Bigl( 1 + {2 c^2 \over k} - uv \Bigr)
\cr}
\eqno(30)
$$

The remaining steps of the procedure for obtaining the solution
are to integrate out the gauge
fields, fix the gauge of the group choosing $b = \pm a$ so that the anomaly
cancels (removing the need for the ``large'' gauge transformation $\Upsilon$
as argued above),
rescale $~\theta' \rightarrow (2c/ \sqrt{k}) ~\theta' ~$ and
take the limit $~c \rightarrow \infty~$ which effectively decouples
the  $~SL(2,R)~$ part from the gauge fields.
The resulting Polyakov sigma model action can
be rewritten as
$$\eqalign{
S_{\sigma ~eff}~=~&-{k \over 8 \pi} \int d^2\sigma
{v^2 \partial_{+}u \partial_{-}u + u^2 \partial_{-}v \partial_{+}v +
(2 - uv)\bigl(\partial_{+}u \partial_{-}v + \partial_{-}u \partial_{+}v \bigr)
\over (1 - uv)} \cr
&+ {q^2 \over 2\pi} \int d^2\sigma \Bigl(
2 (1 - uv) \partial_{+}\theta' \partial_{-}\theta' \Bigr) \cr
&+ {q\sqrt{k} \over 2\sqrt{2}\pi} \int d^2\sigma
\Bigl( \bigl( u\partial_{-}v - v\partial_{-}u \bigr) \partial_{+}\theta' +
\bigl( v\partial_{+}u - u\partial_{+}v \bigr) \partial_{-}\theta' \Bigr) \cr}
\eqno(31)
$$

\noindent A transformation of coordinates
$$
u~=~~e^{\sqrt{2\over k}qt'} \sqrt{{r \over q^2} - 1 }
{}~~~~~~~~~~~~~~
v~=~~-e^{-\sqrt{2\over k}qt'} \sqrt{{r \over q^2} - 1 }
\eqno(32)
$$

\noindent reproduces the solution (25-26).
The dilaton can be found either from the
associated  effective action, as was discussed before,
or from a careful computation of the
Jacobian determinant arising from integrating out the gauge fields [8].
Since the metric-axion solution is exactly (25-26), from the arguments before
the dilaton must be a constant,
$~\Phi = \Phi_0~$. The Jacobian matrix
method confirms this. Its inspection
 before the limit $~c \rightarrow \infty~$ is taken shows
that it is $~J \propto 1/(1 + (2c^2 /k) - uv)~$
$= (k/2c^2)/\bigl(1 + (k/2c^2)(1 - uv)\bigr)~$ (see Ref. [8,17]).
As $~c \rightarrow \infty~$ the non-constant terms decouple and do not
contribute to the dilaton.  Then, the identification procedure can be carried
out along the lines elaborated following Eq. (26).

If one compares the method of obtaining (25-26) to the constructions of Ref.
[1-3], one might object that the
identification has invoked a somewhat arbitrary
step involving the choice of the vector $\xi$ along which the factorization
of $SL(2,R)$ has been performed. The reason for this arbitrariness lies in the
asymptotic properties of the solution (25-26). As $r \rightarrow \infty$, the
metric (25) approaches the vacuum solution with $J=M=0$.
The axion (26) is independent of the mass and spin, and already in the
``vacuum'' form, and so invariant under boosts (13). Therefore, infinitely
far away from the black hole the boost generator is approximately Killing,
and hence the metrics with different spins become locally indistinguishable.
This, on the other hand justifies the interpretation of $J$ as spin, as
discussed in Sec. 3., and the factorization procedure outlined above.
However, it is
possible to generate the spin directly in the effective action of the type of
Eq. (30), and always compactify along the free
boson which appears in the definition of the group. This can be done if one
starts with the group $SL(2,R) \times R^2$, and performs a double gauging$^{*}$
down to a coset $SL(2,R) \times R^2 / R^2$ with
two different vector fields. The subgroups of $SL(2,R) \times SL(2,R)$ are
axial \underbar{and} vector,
mixed with the translations along the two bosons in such way
that the anomaly still cancels. This will necessarily introduce a constraint
on the gauge charges of the two bosons, but it can be satisfied. I hope to
address this issue in a forthcoming paper [18].

{
\small
\baselineskip=12pt plus 1pt
\footnote{}{$^{*}~~$ Double gauging has been employed previously
on different groups in the last of Refs. [7-8] as well as in [11].}
}

\vskip1cm
{\bf 5.~~ Conclusions And Future Interests}
\vskip0.5cm
It is evident that the three dimensional Anti-de-Sitter
black hole of Banados, Teitelboim and Zanelli represents a very
interesting addition to the growing family of black objects. It is
not only a nice example of a black hole in three dimensions, but also
exhibits a surprisingly rich structure. Moreover, it is also a solution
of many different theories of gravity in three dimensions, in growing
order of complexity: General Relativity, Topologically Massive Gravity (TMG),
and String Theory. In this paper, only GR and String Theory have been
explicitly investigated. However, it is not difficult to see that the
solution (12) also represents a solution of TMG with a negative cosmological
constant.

The reason for this is that TMG differs from GR in the presence
of the Lorentz Chern-Simons form in the action. This changes Einstein's
equations by adding the Cotton tensor to the Einstein. Yet both these terms
vanish for the solution (12). This can be seen as follows. The Cotton tensor
is constructed from the covariant derivatives of the Ricci tensor. When the
constructions encountered in this paper are present, the Riemann tensor
is covariantly conserved and so are all other curvature invariants. Hence the
Cotton tensor is zero. Furthermore, since by the boost (13) the solution (12)
can be brought locally in the diagonal form (15), and the Lorentz Chern-Simons
form of (15) is trivially zero, it also vanishes for (12). Therefore,
as claimed, the solution (12) also solves the equations of motion of TMG
in a trivial way.

One other interesting feature of (12) is that it can be used for constructing
a black string configuration in four dimensions. Effectively, all one needs to
do is to tensor (12) with a flat direction $R$. Related considerations
have been investigated previously in [5]. There remains a subtlety regarding
the interpretation of such solutions. I hope to return to this question
elsewhere [18].

In summary, the three dimensional black hole has shown great promise for
surprise. It is a truly multifaceted configuration, which possesses rich
geometrical structure, and appears to relate different formulations of gravity
by being a solution of all of them. One can not resist the temptation, that
perhaps this is not an accident, but rather a beacon pointing at a certain,
more fundamental, interconnectedness of these theories of gravity.

\vskip1cm
\noindent {\bf Note added in proof:} Upon the completion
of this work I have received
Ref. [19] which overlaps in some length with the subject of this work, and
where similar results were found.

\vskip1cm
{\bf Acknowledgements}
\vskip0.5cm
I have benefited immensely from conversations with
B. Campbell, G. Hayward
and D. Page. I would also like to thank  W. Israel
for pointing out Ref. [14] to me.
This work has been supported in part by
the Natural Science and Engineering Research Council of Canada.

\vskip1cm
{\bf References}
\vskip0.5cm
\item{[1]} M. Banados, C. Teitelboim, and J. Zanelli, Phys. Rev. Lett.
{\bf 69} (1992) 1849.
\item{[2]}M. Banados,  M. Henneaux, C. Teitelboim and J. Zanelli, IAS preprint
HEP-92/81 /gr-qc/9302012, to appear in Phys. Rev {\bf D}.
\item{[3]} D. Cangemi, M. Leblanc and R.B. Mann,
MIT preprint CTP-2162, 1992/gr-qc/9211013.
\item{[4]}S.F. Ross and R.B. Mann,
Univ. of Waterloo preprint WATPHYS TH 92/07, Aug. 1992/hep-th/9208036.
\item{[5]} N. Kaloper, Univ. of Minnesota preprint UMN-TH-1024/92, May 1992,
in press in Phys. Rev. {\bf D}.
\item{[6]} E. Witten, Phys. Rev. {\bf D44} (1991) 314.
\item{[7]} N. Ishibashi, M. Li and A.R. Steif, Phys. Rev. Lett. {\bf 67}
(1991) 3336; P. Ginsparg and F. Quevedo, Nucl. Phys. {\bf B385} (1992) 527;
I. Bars and K. Sfetsos, USC preprint USC-92/HEP-B1, May 1992/hep-th/9301047;
Phys. Rev. {\bf D46} (1992) 4510;
K. Sfetsos, USC preprint USC-92/HEP-S1, June 1992; I. Bars and K. Sfetsos,
USC preprint USC-92-HEP-B3, Aug. 1992/hep-th/9208001.
\item{[8]} P. Horava, Phys. Lett. {\bf B278} (1992) 101;
E.B. Kiritisis, Mod. Phys. Lett. {\bf A6} (1991) 2871;
D. Gershon, Tel Aviv University preprint TAUP-1937-91, Dec 1991/hep-th/9202005.
\item{[9]} J.H. Horne and G.T. Horowitz, Nucl. Phys. {\bf B368} (1992) 444.
\item{[10]} E. Raiten, Fermilab preprint FERMILAB-PUB-91-338-T, Dec. 1991.
\item{[11]} C.R. Nappi and E. Witten, Phys. Lett. {\bf B293} (1992) 309.
\item{[12]} E. Cremmer, ``Dimensional Reduction in Field Theory and Hidden
Symmetries in Extended Supergravity'', pp 313, in {\bf Supergravity 81},
ed. S. Ferrara and J.G. Taylor, Cambridge Univ. Press, Cambridge 1982.
\item{[13]} S. Deser, R. Jackiw and S. Templeton,
Ann. Phys. {\bf 140} (1982) 372.
\item{[14]} C. Misner, ``Taub-NUT Space as a Counterexample to Almost
Anything'', in {\bf Relativity Theory and Astrophysics}, ed. J. Ehlers, Am.
Math. Soc., Providence RI 1967.
\item{[15]} J. Antoniades, C. Bachas, J. Ellis and D. Nanopoulos,
Phys. Lett. {\bf B221} (1988) {393; J. Antoniades, C. Bachas and A. Sagnotti,
Phys. Let. {\bf B235} (1990) 255.
\item{[16]} for a review, see L. Alvarez-Gaume and M.A. Vasquez-Mozo,
``Topics on String Theory and Quantum Gravity'' Les Houches Lectures,
CERN preprint CERN-TH-6736-92, Nov. 1992/hep-th/9212006.
\item{[17]} N. Kaloper, Mod. Phys. Lett. {\bf A8} (1993) 421.
\item{[18]} N. Kaloper, in preparation.
\item{[19]} G.T. Horowitz and D.L. Welch,
UCSB preprint NSF-ITP-93-21/hep-th/9302126

\bye